\documentclass[a4paper,11pt]{article}

\usepackage{jcappub} 

\usepackage[T1]{fontenc} 
\usepackage{epstopdf}
\def\beq{\begin{eqnarray}}                      
\def\eeq{\end{eqnarray}}                         
\def\nn{\nonumber}                                 
\def\n{\label}                                           
\def\r{\ref}                                              

\definecolor{linkcolor}{rgb}{0.0,0.3,0.5}
\definecolor{venetianred}{rgb}{0.78, 0.03, 0.08}

\def\beq{\begin{eqnarray}}
\def\eeq{\end{eqnarray}}
\def\ln{\,\mbox{ln}\,}

\def\al{\alpha}

\def\ga{\gamma}

\def\la{\lambda}
\def\na{\nabla}

\def\si{\sigma}

\def\ph{\varphi}

\title{\boldmath On higher derivative
corrections to the
\texorpdfstring{$R+R^2$}{Lg}
inflationary model}
\author[a]{Ana R. Romero Castellanos,}
\author[a]{Flavia Sobreira,}
\author[b,c,d]{Ilya L. Shapiro,}
\author[e,f]{Alexei A. Starobinsky}
\affiliation[a]{Instituto de F\'{\i}sica Gleb Wataghin, Universidade Estadual de Campinas,\\13083-859, Campinas,  SP, Brazil}
\affiliation[b]{Departamento de F\'{\i}sica, ICE, Universidade Federal de Juiz de Fora,\\ 36036-330, Juiz de Fora, MG, Brazil}
\affiliation[c]{Tomsk State Pedagogical University,\\ Tomsk, 634041, Russian Federation}
\affiliation[d]{Tomsk State University,\\ Tomsk, 634050, Russian Federation}
\affiliation[e]{L. D. Landau Institute for Theoretical Physics RAS,\\Moscow, 119334, Russian Federation}
\affiliation[f]{Kazan Federal University,\\Kazan 420008, Republic of Tatarstan, Russian Federation}
\emailAdd{arromero@ifi.unicamp.br}
\emailAdd{sobreira@ifi.unicamp.br}
\emailAdd{shapiro@fisica.ufjf.br}
\emailAdd{alstar@landau.ac.ru}

\abstract{The $R+R^2$ model is successful in describing inflation, as it provides an excellent fit to the full set of available observational data. On the other hand, the same model is the simplest extension of general relativity which does not produce higher derivative ghosts and related instabilities. Long ago, it was proposed to treat all terms which cause higher derivative instabilities as small perturbations that could avoid the presence
of ghosts in the spectrum. We put this proposal into practice and consider an explicit example of treating more complicated higher derivative terms as small perturbations over the $R+R^2$ model by introducing the $R\Box R$ term into the action. Within the described scheme, it is possible to obtain an upper bound
on the coefficient of this non-scale-free sixth-derivative term by mapping the theory into a one-scalar field potential. It is shown that the result differs from treating this term on equal footing with other terms that requires mapping to a two-scalar field model, and in general leads to different observational consequences.
}
\begin{document}
\maketitle
\flushbottom
\section{Introduction}
The role of fourth order higher derivative gravitational terms can be seen from
different perspectives. In quantum field theory in curved space-time and in the
semiclassical approach to gravity, these terms are required to provide a
renormalizable theory and to obtain a finite renormalized average value of
the energy-momentum tensor of non-gravitational quantum fields
\cite{Utiyama:1962sn} (in cosmology this was first done in
\cite{Zeldovich:1971mw}, see \cite{Shapiro:2008sf} for a review and
further references). The same situation
holds in quantum gravity, where these terms provide renormalizability
\cite{Stelle:1976gc}. An important advantage of higher derivative
terms is that their effects are strongly suppressed below the Planck scale,
and hence the classical solutions of general relativity can be seen
as a very good approximation at the cosmological, astrophysical
and laboratory scales. For this reason, theories with high derivative
terms do not contradict experimental and observational tests of
Einstein's gravity.

On the other hand, these terms lead to the
violation of stability of the classical solutions. At the quantum
level, one can find violation of unitarity due to the presence of
states with negative energy or negative norm \cite{Stelle:1976gc}.
No comprehensive solution to this problem is known, but the
main expectations were always posed on the theories with complex
poles, which were supposed to emerge due to loop contributions
\cite{Salam:1978fd,Tomboulis:1977jk,Tomboulis:1980bs,Tomboulis:1983sw,Antoniadis:1986tu}. Indeed, our knowledge of quantum corrections to the gravitational
propagator is insufficient to know whether the solution of the ghost
problem can be achieved in this way \cite{Johnston:1987ue}. At the
same time, there are versions of super-renormalizable quantum gravity
\cite{Asorey:1996hz}, which have complex conjugate poles already
at the classical level, and in this case one can prove unitarity in the
Lee-Wick sense \cite{Modesto:2015ozb}. These models have sixth
or higher order derivatives in the action, and represent a prospective
subject of investigation, in particular in cosmology.

At low energies, the terms with higher derivatives can be regarded
as small perturbations of the fiducial theory of general relativity.
This approach has been suggested as an {\it ad hoc} universal
solution to the ghost problem \cite{Simon:1990ic}. This proposal
leads to the following dilemma: trying to consider {\it all} higher
derivative terms as objects to be avoided at the fundamental level
and treated as perturbations, one has to 'forbid' the $R+R^2$ (Starobinsky)
model of inflation -- the simplified variant of the model introduced in
\cite{Starobinsky:1980te} -- arguing that it is 'non-perturbative'
\cite{Simon:1991bm} (and it does, in fact, though it is not
non-perturbative with respect to general scalar-tensor gravity). Indeed,
this part of the proposal is something difficult to accomplish. First
of all, the  $R^2$-term does not produce ghosts and hence there is
no reason to avoid it. At the same time, this inflationary model is
the most successful from the observational and phenomenological
point of view, so it is not easy to give it up without a real
motivation. Finally, the inflation scenario requires the value of
the numerical coefficient of the $R^2$-term to be quite big,
about $5 \times 10^8$ \cite{Starobinsky:1983zz} (see also the recent
papers \cite{kaneda2010fourth,Netto:2015cba}).
This makes this term dominant at curvatures much less than the Planck
one, in particular, at curvatures during inflation. As a result, it would be
quite natural to replace the $R^2$-term into another side of the perturbation
scheme of \cite{Simon:1990ic} and include it into the basic action along
with the Einstein-Hilbert term. The present work is devoted to the
practical application of this idea to inflation. Namely, we add a
small sixth-derivative term to the standard inflationary $R+R^2$
action and find an upper bound on the coefficient of the new term,
treating it as a small perturbation.

Previous attempts to analyze higher derivative corrections to the
Einstein-Hilbert actions have been performed, e.g., in
\cite{Gottlober:1989ww,berkin1990effects,Amendola:1993bg,chiba2005generalized,Cuzinatto:2013pva}, making a
transformation of the gravitational terms to scalar fields, among those
one or more ghosts appear, or by considering non-local generalizations
of gravity which does not contain ghosts
\cite{craps2014cosmological,koshelev2016occurrence} and in which
solutions of a local higher derivative theory like the $R+R^2$ one can
appear as exact particular solutions of non-local equations.\footnote{When
this manuscript was prepared for publication, a new paper on this topic
\cite{Carloni:2018yoz} appeared which belongs to the former class.} Let
us also mention an earlier work \cite{Maroto:1997aw}, where the dS-type
solutions were explored in the framework
of string-induced fourth-derivative gravity. In the present paper, we
follow a different approach and treat the term which may produce
ghosts as a small perturbation, such that it becomes harmless.
In what follows one can find the perturbative analysis for a special
sixth-order term which leads to the constraints on its coefficient
derived from observational predictions.

Let us give a comment on the choice of the particular form of the
term which will be used below to represent higher derivatives. Since
all such terms are supposed to be small perturbations, one can
consider them one by one, and in the leading-order approximation
the effects of these terms will not depend on each other. As a
consequence, we can actually start with an especially simple example
by considering the $R \square R$-term. As we shall see in what
follows, this term is simple to deal with, and gives a clear idea of
a way in which one can consider generic higher derivative terms as
perturbations. Finally, since the present cosmological constant does not play
any noticeable role in the $R+R^2$ inflationary model, we set it to zero.

The paper is organized as follows. In Sec.~\ref{s2} we review the
standard mapping of $R+R^2$ gravity into a metric-scalar model. In
Sec.~\ref{s3} it is shown how an extra $R \square R$-term can be
introduced as a small perturbation of this model, analyzing the
observational consequences of the theory in the slow-roll regime.
It is shown that when the $R \square R$-term is treated as a small
perturbation, the model can still be mapped into a {\it one-scalar theory},
different from the {\it two-scalar} mapping within the approach
which is a standard procedure in cosmology. The approach
with the mapping to the two-scalar model is considered in parallel,
for the sake of comparison, and the well-known necessary details
of this presentation are postponed to the Appendix. Finally, the
reader can see that when the $R \square R$-term is treated at the
same level with other terms,  it is mapped into the model with
{\it two} scalars, and the observational consequences are in
general different. Finally, in Sec.~\ref{s5} we draw our conclusions.

\section{The \texorpdfstring{$R+R^2$}{Lg} Model}
\label{s2}
Among different models of inflation
\cite{linde1984inflationary, olive1990inflation, bassett2006inflation},
the $R+R^2$ model introduced in \cite{Starobinsky:1980te} is one
of the most appealing from both theoretical and observational
perspectives. It has the least number (one) of free parameters fixed by
observations only. The action of this model is closely related to vacuum
quantum corrections \cite{Zeldovich:1971mw,Fischetti:1979ue} (see also
\cite{Fabris:2000gz,Fabris:2003gp} and \cite{Netto:2015cba} for
the recent advances in this direction) and, on the other hand, its
predictions are consistent with recent bounds including the ones set
by the Planck collaboration \cite{Akrami:2018vks,Akrami:2018odb,aghanim2018planck}.

The model is described by the Einstein-Hilbert action with an extra
term proportional to the square of the Ricci scalar $R$,
\beq
\label{Action0}
S_0=\frac{M_{P}^2}{2}\int d^4x \sqrt{-g}\,
\big(R+\alpha R^2 \big),
\eeq
where $M_{P}$ is the reduced Planck mass, $\alpha=(6M^2)^{-1}$
where $M$ is the low-curvature ($|R|\ll M^2$) value of the rest mass of the
scalar degree of freedom (dubbed scalaron in \cite{Starobinsky:1980te})
appearing in $f(R)$ gravity, and we put $\hbar = c =1$. The theory
 (\ref{Action0}) can be easily mapped into a metric-scalar model (see, e.g.,
\cite{PhysRevD.83.084028} and \cite{huang2014polynomial} where the
procedure is described for the general $f(R)$ extension)
\beq
\label{ActionPhi}
S^*_0 =\frac{M_{P}^2}{2} \int d^4x \sqrt{-g}\,
\big[\phi_0 R-U_0(\phi_0) \big],
\eeq
where the scalar field $\phi_0$ is related to the Ricci scalar by the relation
\beq
\phi_0=1+2\al R,
\label{ph0}
\eeq
and the potential function $U_0(\phi_0)$ is
\beq
 U_0(\phi_0)=\frac{1}{4\al}(1-\phi_0)^2.
\eeq
It proves useful to make a conformal transformation,
introducing a new scalar field $\chi_0$,
\beq
\bar{g}_{\mu\nu}
&=&
g_{\mu\nu}\,\exp \left\{
\textstyle{ \sqrt{\frac23}\,\frac{\chi_0}{M_P}}\right\}.
\eeq
The action which results from this procedure has a standard
kinetic term, and reads
\beq
\label{actionS}
S^*_0=\int d^4x\sqrt{-\bar{g}}
 \Big[
 \frac{M_{P}^2}{2}\bar{R}
 - \frac{1}{2} (\bar{\na}\chi_0)^2 - V(\chi_0)\Big],
\eeq
where $\,(\bar{\na}\chi_0)^2 =
\bar{g}^{\mu\nu}(\bar{\na}_{\mu}\chi_0)(\bar{\na}_{\nu}\chi_0)\,$
and $\,V(\chi_0)\,$ is a potential given by the expression
\beq
\label{Potencialcero}
 V(\chi_0)
\, =\,
 \frac{M_{P}^2}{8\al}
 \left(1-e^{-\frac{2}{\sqrt{6}}\frac{\chi_0}{M_{P}}}\right)^2,
\eeq
which drives the evolution of the scalar field $\chi_0$ (scalaron) and satisfies
the slow roll conditions in the large field regime.

\section{\texorpdfstring{Treating $R \Box R$}{Lg} term
as a small perturbation to the \texorpdfstring{$R+R^2$}{Lg} model}
\label{s3}

Let us now consider the modification of the scheme described above
when introducing an extra term $R\Box R$ treated as a perturbation. We
shall start from a brief review of the previously known way of dealing
with this term, while technical details can be found in the Appendix. It is
expected that the comparison of the two approaches will make their differences
clear.

The new action is
\begin{equation}
\label{R-action}
 S=\frac{M_P^2}{2}\int d^4x\sqrt{-g}\left[R+\alpha R^2
 +\gamma R\Box R \right],
\end{equation}
where the parameters $\al$ and $\ga$ have dimensions of
$[\textrm{mass}]^{-2}$ and $[\textrm{mass}]^{-4}$, respectively.
The new term is the simplest one leading to a ghost, as described
in \cite{Accioly:2016qeb}, and therefore it is interesting to see how
it can be treated as a small perturbation, while the ghost problem
is avoided.

The action (\ref{R-action}) can be written in terms of two scalar
fields \cite{Gottlober:1989ww,chiba2005generalized} (see  Appendix
\ref{ap1} for details):
\beq
\label{sfinal}
 S
 &=&
 \frac{M_P^2}{2}\int d^4x\sqrt{-\bar{g}}\big[
 \bar{R} - 6({\bar{\nabla}\varphi})^2-\gamma e^{-2\varphi}
 (\bar{\nabla}\phi_1)^2
 \nonumber
 \\
 & - &
 U(\phi_1,\varphi)\big],
\end{eqnarray}
where the potential is defined as
\begin{equation}
\label{potential}
 U(\phi_1,\varphi)=e^{-4\varphi}\left[\phi_1(e^{2\varphi}-1)-\alpha\phi_1^2\right].
\end{equation}
From Eq. (\ref{action1_apen}), it can be seen that in this case the fields $\varphi$ and $\phi_1$ are related with the Ricci scalar $R$ and
$\Box R$ by
\begin{eqnarray}
 \varphi&=&\frac{1}{2}\ln (1+2\alpha R +2\gamma \Box R),\nonumber\\
 \phi_1&=&R,
\end{eqnarray}
that is in complete agreement with the results presented in \cite{Gottlober:1989ww}.
From equations (\ref{sfinal}) and (\ref{potential}), it is clear
that when $\gamma=0$ we recover the  $R+\alpha R^2$ case
\cite{Starobinsky:1980te} in the Einstein frame.
In appendix \ref{ap1} we present the derivations of the Einstein
equation in the weak field approximation up to the second order in
the fields that reproduces the results of \cite{Gottlober:1989ww}.
Also, the Ansatz $\Box R= r_1 R+r_2$ proposed by
\cite{craps2014cosmological}, is used to verify that the results are
consistent with the ones presented in \cite{koshelev2016occurrence}.
Indeed, the action (\ref{sfinal}), which follows from the
standard approach, has a non-standard kinetic term, of the type that
were analyzed in several works
~\cite{mukhanov1998density,garcia1995constraints,garcia1996metric,
di2005slow}, always in the slow roll approximation.

Nevertheless, this kind of analysis is not completely free of problems,
because equations following from the action (\ref{sfinal}) does not satisfy
the slow roll conditions generically, as discussed in \cite{wang2010note}.
In this situation, definition of new slow roll parameters and alternative
treatments have been proposed, for instance,
in ~\cite{ji2009curvature,lalak2007curvature,wang2018noncanonical}.

However, the problem is that the ``standard'' treatment of the situation in
cosmology which was done in the references mentioned above
is opposite to the one which is usually considered ``standard'' in
dealing with higher derivative theories \cite{Simon:1990ic}. Thus, we
will close this gap and consider the last term in Eq.~(\r{R-action})
as a perturbation.  The main point is that then we cannot use the
standard scheme of mapping to the metric-scalar models (see, e.g.,
\cite{PhysRevD.83.084028}). Instead, we have to follow the
treatment of the new term as a perturbation that means that the number
of degrees of freedom is not increased in contrast to the action
Eq.~(\ref{sfinal}).

Treating the $R\Box R$-term as a perturbation, one can
suppose that in mapping to a scalar-metric model, the term
$R\Box R$ should be substituted by $R(\phi_0)\Box R(\phi_0)$,
where $\phi_0$ is a scalar field, similar to $\phi_0$ in
Eq.~(\ref{ActionPhi}). The action is perturbed by the inclusion
of the $\ga$ term,
\beq
 S \,=\, S^*_0+S_{\ga},
\eeq
where $S_{\ga}$ is defined as
\beq
 S_{\gamma}
 &=&
  \frac{M_{P}^2\ga}{2}\int d^4x \sqrt{-g}
 R\Box R \Big\arrowvert_{R=\frac{(\phi_0-1)}{2\al}}
\nonumber
 \\
 &=&
  \frac{M_{P}^2\ga}{8\al^2} \int d^4x \sqrt{-g} \phi_0\Box \phi_0.
\eeq
Let us note that the relation between $R$ and $\phi_0$ in this formula
is exactly the same as the obtained for the
unperturbed $R+R^2$ model given by equation (\ref{ph0}) without
any changes. This procedure means that we disregard all possible
terms of higher orders in $\ga$.

In order to obtain the scalar mapping under this approximation, we
use the relation between the $\si=e^{2\ph}$ and $\phi_1$ fields
given by the equation of motion (\ref{eqmovphi1sigma}), taken up to
linear order in the  $\frac{\ga}{2\al}\Box$ operator. This leads to
\beq
\label{appphi1}
\phi_1 &=&
\frac{1}{2\al}(e^{2\ph}-1)-\frac{\ga}{2\al^2} \Box e^{2\ph}.
\eeq
When Eq. (\ref{appphi1}) is substituted back into the action, this gives us
\beq
&& S
=
\frac{M_P^2}{2}
\int d^4x \sqrt{-\bar{g}}
 \Big[\bar{R}-6({\bar{\nabla}\varphi})^2
 \big(1+ke^{2\ph} \big)
- U(\ph)\Big],
\nn
\\
&&
\mbox{where}
\quad
 k = \frac{\gamma}{6\alpha^2}
\quad
\mbox{and}
\quad
U(\ph)=\frac{e^{-4\ph}}{4\al}(e^{2\ph}-1)^2
\label{approx_action}
\eeq
which has exactly the  same form as the one in Eq.~(\ref{pot_ansatz}) in
the Appendix. Nevertheless, in this case the kinetic term has
a non-canonical form. Thus, to find the mass of the field in the Minkowski limit,
it has to be transformed into the standard form.
It is important to emphasize that all the analysis made in this work are concerned with the
case $|k|\ll 1$, in order to allow the treatment of the $\Box R$ term as a small perturbation to
the $R+R^2$ theory.
It is easy to see that when the Ansatz (\ref{ansatz}) is used
alongside with the previous approximation, we recover the results
of \cite{craps2014cosmological,koshelev2016occurrence}. In order
to check this, we write the action (\ref{approx_action}) in the
equivalent way,
\beq
\label{Senlightening}
S
&=&
\frac{M_P^2}{2}\int d^4x\sqrt{-\bar{g}}
\Big[
\bar{R}-6({\bar{\nabla}\varphi})^2
+ 3k(\sigma-1)\bar{\Box}\ph
\nn
\\
& - &U(\ph)\Big],
\quad
\mbox{where}
\quad
\si=e^{2\varphi},
\eeq
as defined above.
Now, we can use Eq.~(\ref{appphi1}) to find the relation
\beq
\left(1-\frac{\gamma}{\alpha}\Box\right)^{-1}
\left[r_1+\Box\si - \frac{\ga}{\al}\Box(\Box\si)\right]=r_1 \si.
\eeq
When expanded up to the linear order in the operator
$\textstyle{\frac{\ga}{\al}}\Box$ we arrive at the relation
$\Box\sigma=r_1(\sigma-1)$.

With these considerations, the action (\ref{Senlightening}) becomes
\beq
S=\frac{M_P^2}{2}\int d^4x\sqrt{-\bar{g}}
\left[\bar{R}-6({\bar{\nabla}\varphi})^2-\mathcal{U}(\ph)\right],
\eeq
where the potential is defined  as
\beq
\mathcal{U}(\ph)=\frac{\al-r_1\ga}{4\al^2}e^{-4\ph}(1-e^{2\ph})^2.
\eeq
Comparing the last expression with the potential corresponding to the
nonperturbed Starobinsky model in the Einstein frame given by
Eq.~(\ref{Potencialcero}), one can see
that the values of the spectral index $n_s$ and the tensor-to-scalar ratio
$r$ will not be modified that is consistent with the results for these
quantities in \cite{koshelev2016occurrence} which are based on the use
of the specific Ansatz (\ref{ansatz}) .

Let us stress that although the term $\ga R\Box R$
studied here is a particular case of the $F(\Box)$ analyzed in
previous works, in our case the simplifying Ansatz (\ref{ansatz}) is not an
additional general requirement but only a special case. Thus, our study in
this paper is not a particular case of that in
\cite{craps2014cosmological,koshelev2016occurrence}, but represents a
qualitatively new approach to the $R\Box R$ term in the action.

The bilinear part of the action (\r{Senlightening}) can be cast
into the form
\beq
 S
 &=&
\int d^4x\sqrt{-\bar{g}}
\Big\{
\frac{M_P^2}{2}\bar{R} - \frac{1}{2}({\bar{\nabla}\chi})^2
- \frac{1}{2}M_{\chi}^2\chi^2\Big\},
\label{Mod}
\eeq
where fields $\chi$ and $\ph$ are related by the relation
\beq
\ph=\frac{\chi}{\sqrt{6\left(1+\ga/6\al^2\right)M_P}}.
\eeq
Let $k\equiv\frac{\ga}{6\al^2}$. If $\left\vert k\right\vert \ll 1$,
the mass of the field $\chi$ is
\beq
\label{corrmass}
 M_{\chi}^2\approx M^2-\frac{\ga}{36\al^3},
\eeq
where $M^2$ (defined below Eq.~(\ref{Action0})) is the scalaron
mass in the Starobinsky model. From Eq.~(\ref{corrmass}) one can
see that the mass of the scalar $\chi$ can be greater or smaller than $M$
depending on the sign of the parameter $\gamma$. This behavior is
preserved by small values of $\left\vert k\right\vert \ll 1$
in the general case of (\ref{approx_action}), as it will be seen
from the estimated values obtained using the Planck results
\cite{aghanim2018planck}.

\begin{figure}
\centering

\includegraphics[scale=1.0]{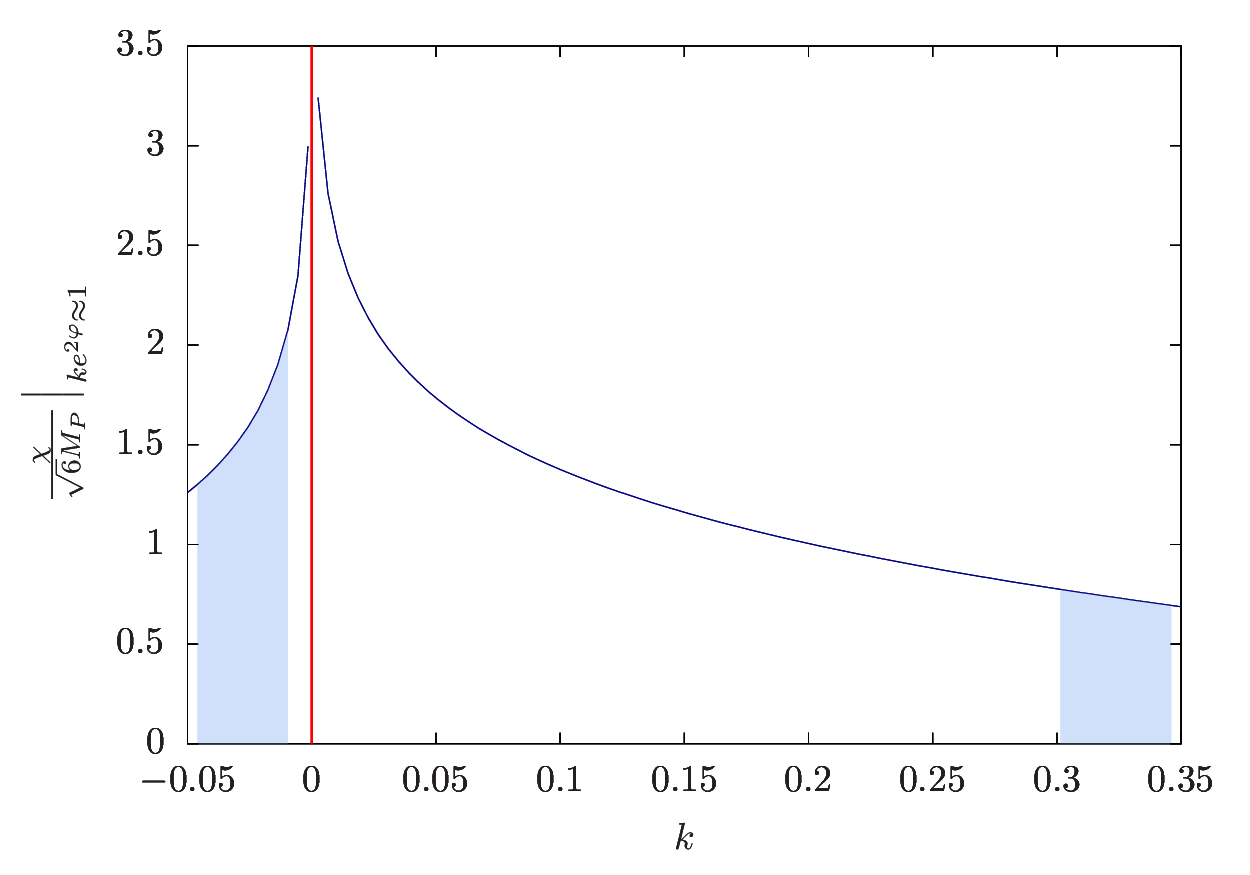}
\caption{The energy scale of the new scalaron $\chi$ as a function
of the perturbative parameter $k$. The shaded regions are excluded
by the constraints imposed by Planck data\cite{aghanim2018planck}, as
will be shown later.}\label{chispecial}
\end{figure}

Returning to the general expression for the action
(\ref{approx_action}), the field transformation that turns the
kinetic term in a canonical form should satisfy
\beq
\left( \frac{d \chi}{d \ph}\right)^2
\,=\,6M_P^2 \left( 1 + k e^{2\ph}\right),
\label{Eq k}
\eeq
and then the action becomes
\beq
S=\int d^4x\sqrt{-\bar{g}}
\Big\{\frac{M_P^2}{2}\bar{R}-\frac{1}{2}({\bar{\nabla}\chi})^2
-V(\ph(\chi))\Big\}.
\eeq
In the Einstein frame, the potential is given by
\beq
\label{corrpot}
V(\chi) \,=\,V(\varphi(\chi)) \,=\,
 \frac{M_P^2}{8\alpha}\left(1-e^{-2\varphi(\chi)}\right)^2,
\eeq
where the dependence of the intermediate field $\ph$ on the scalar field
$\chi$ in Eq.~(\ref{corrpot}) (which may be called new scalaron) can be
obtained by solving the transcendental equation that follows from (\ref{Eq k}),
\beq
\label{rel_chi_phi}
\frac{\chi}{\sqrt{6}M_P}
&=&
\ln\left[\frac{1-\sqrt{1+ke^{2\ph}}}{e^{\ph}(1-\sqrt{1+k})} \right]+ \sqrt{1+ke^{2\ph}}-\sqrt{1+k}.
\eeq
In the last expression for the particular case $ke^{2\varphi}\approx 1$,
and taking into account that $|k|\ll 1$, but $k\neq 0$ we have
\begin{equation}
\label{eqchispecial}
\left.\frac{\chi}{\sqrt{6}M_P}\right|_{ke^{2\varphi}\approx 1}\approx -\frac{1}{2}\ln{|k|}-\frac{k}{4}+\frac{1}{4}\textrm{Sign}(k),
\end{equation}
which shows that a large field inflation can be developed when
$|k|\ll1$, as can be seen from figure \ref{chispecial}, where we
have plotted Eq. (\ref{eqchispecial}) which shows the behavior
of the new scalaron $\chi$ in the particular case
$ke^{2\varphi}\approx 1$, where the shaded regions are excluded
by the Planck data, as will be discussed below. 

Let us note
that the condition $ke^{2\varphi}\approx 1$ has been chosen since
it corresponds to the maximal value of the parameter $k$ which is
compatible with our approximation $\vert k\vert \ll 1$. For smaller
values of $\vert k\vert$ the effect of the $R\Box R$ term will be
less significant.

Let us note that the general expression (\ref{rel_chi_phi}) can be
used as a basis of metric-scalar cosmological model even for the
values of $k$ which do not satisfy the condition $|k|\ll 1$. However
in this case there is no direct link with the main perturbative in
$ R \Box R$ approach for the  $R+R^2 + R \Box R$  theory,
which we aim to develop in this work.

\begin{figure}[htb]
\centering
\includegraphics[scale=0.9]{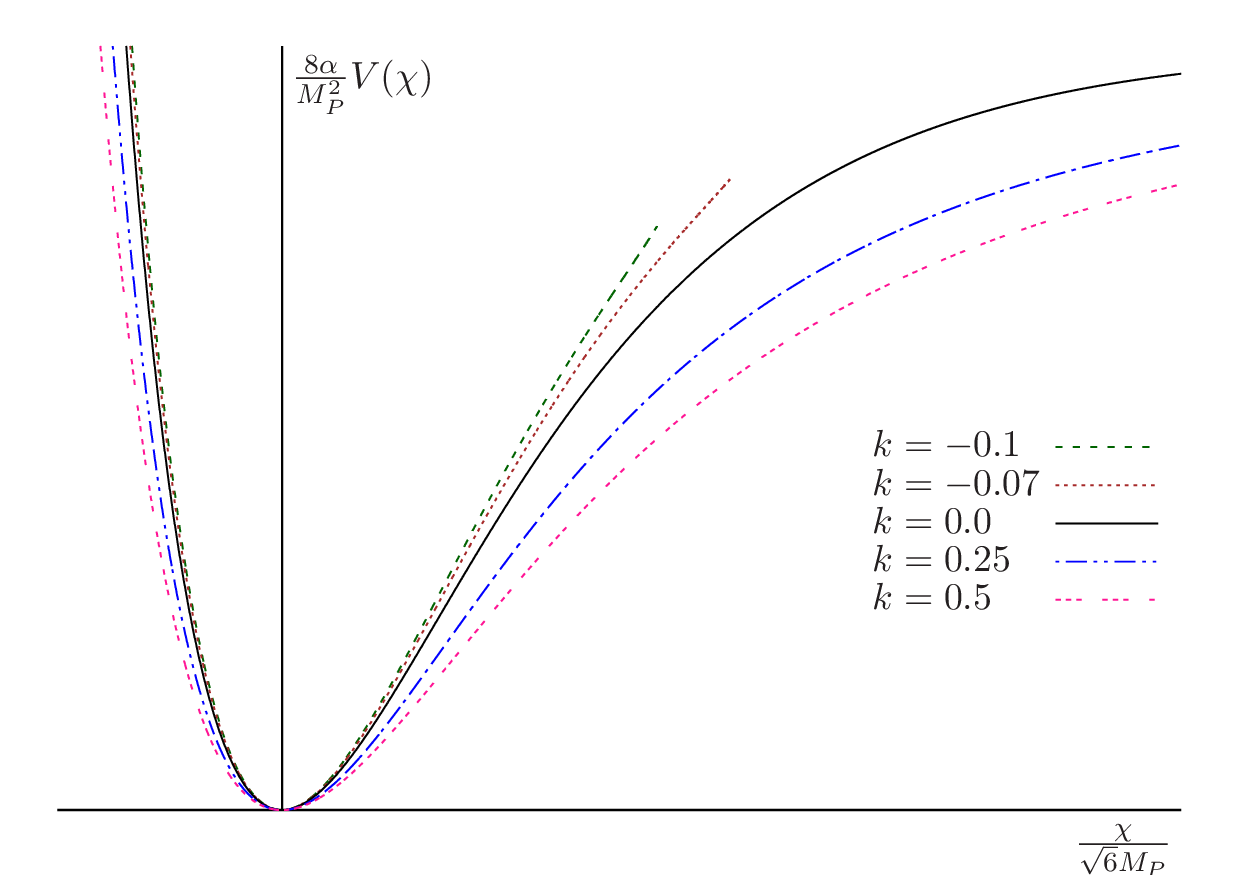}
\caption{$R\Box R$ correction to the Starobinsky model of inflation
for different values of the parameter $k$. The black continuous
line corresponds to the original $R+\al R^2$ model, and the
colored dashed lines to nonzero values of $k$.}
\label{graphpotscalar}
\end{figure}

In Fig.~\ref{graphpotscalar}, the plot of the potential $V(\chi)$ is
shown for different values of the parameter $k$, with $k=0$
corresponding to  the $R+R^2$ model, and the other values to the
extra $R \Box R$ term with different coefficients. One can observe that
the presence of the $R\Box R$ term changes the shape of the potential
including its flatness. For $k > 0$, the slow-roll inflationary regime ends for
slightly larger values of the scalar field, implying that for larger
values of $k$, inflation happens at higher energy scales than
for the standard $R+R^2$ model.

For positive $k$, all expanding spatially-flat FRLW universes evolve
to the dust-like one filled by massive scalarons at rest at late times,
like in the case of the Starobinsky model. As we will see later, at
earlier times, they can develop inflation in the slow roll regime. For
the case of negative values of $k$ (remember $|k|\ll 1$), there is a
maximum in the potential for a critical value of the $\chi$ field, given
by the expression
\begin{equation}
\frac{\chi_{\rm max}}{\sqrt{6}M_P}=\ln\left(\frac{\sqrt{|k|}}{1-\sqrt{1-|k|}}\right)-\sqrt{1-|k|},
\end{equation}
which comes from the constraint of a real $\chi$ field in Eq. (\ref{rel_chi_phi}).
In this case, slow roll inflation can take place only for values of $k$ which are
 close to zero, as will be seen from the behavior of the slow roll parameters
$\epsilon$ and $\eta$. In fact, this is not too relevant, since all our analysis is
valid only for small values of the parameter $|k|$.

The nonzero $k$ or, equivalently, the $\ga$ term, modifies the value of $\chi$ in
which the last 60 $e$-folds of inflation begin, leading to a modification of
observable parameters such as the tilt of the primordial power spectrum of scalar
(adiabatic) metric perturbations $n_s$  and the scalar-to-tensor ratio $r$, as we
will see in below. Furthermore, the  $R\Box R$-type perturbation modifies the
symmetry of the scalar potential near its minimum that can affect oscillations of
the field $\chi$ (new scalaron) after inflation and gravitational creation of
particles and antiparticles by these oscillations through parametric resonance
\cite{Starobinsky:1980te,Starobinsky:1981vz,kofman1994reheating,kofman1997towards}.
The next important question is whether a non-zero  $\ga$
modifies the conditions of a slow roll inflation.

\subsection{Slow-roll conditions}

As far as the model with non-zero $\ga$-term is mapped into a
single scalar field action, the analysis of the slow roll conditions
can be performed in a standard way.

\begin{figure}
\centering
\includegraphics[scale=0.85]{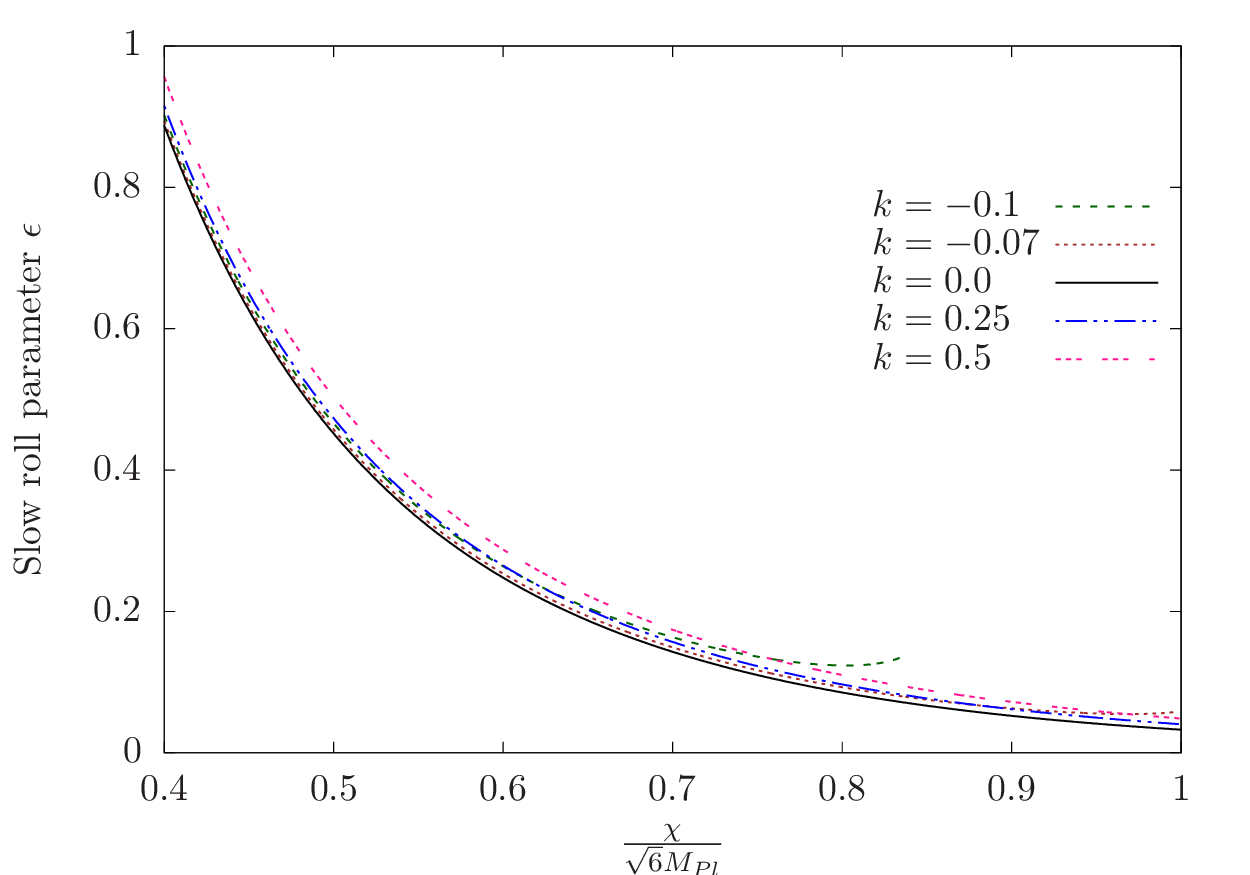}
\includegraphics[scale=0.85]{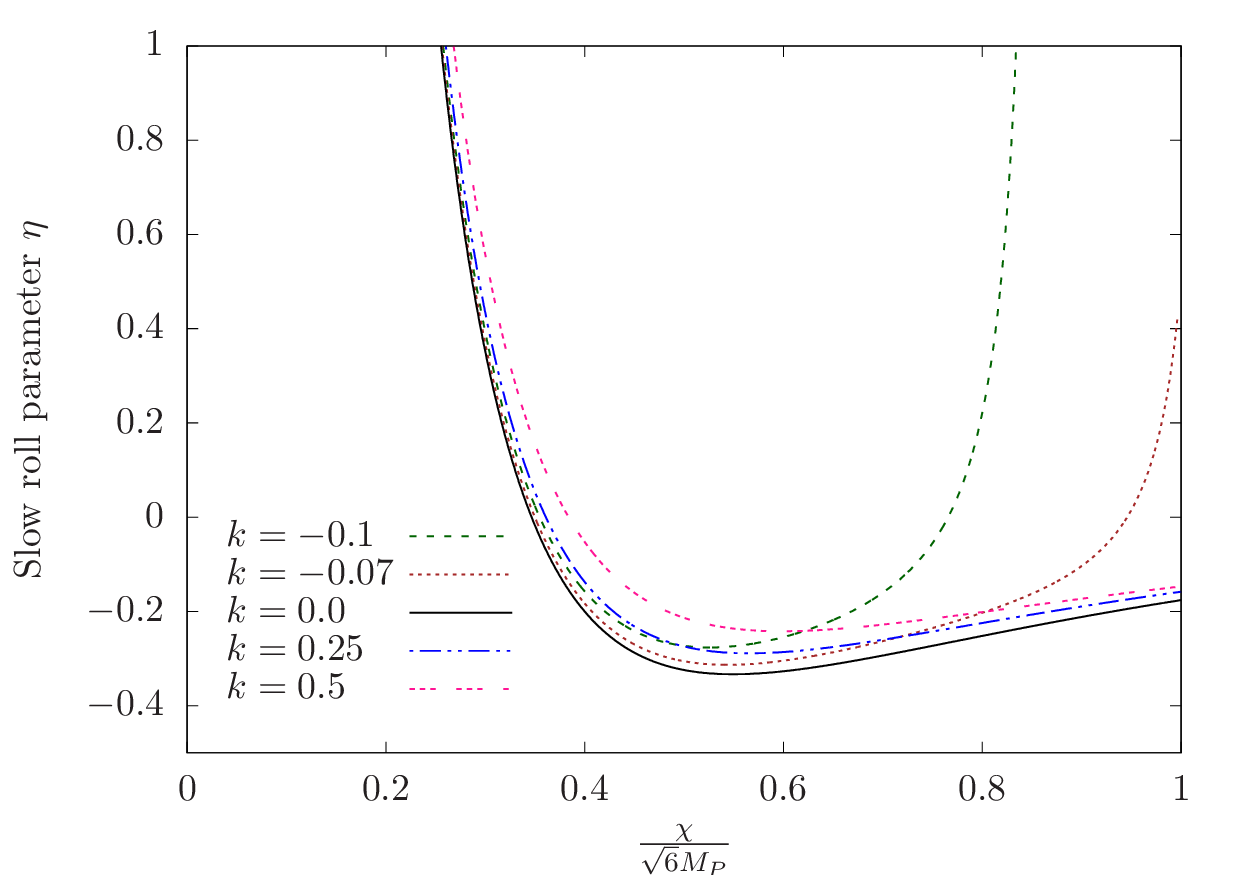}
\caption{The slow roll parameters $\epsilon$ (up) and $\eta$ (down)
for the model with $R\Box R$ term as functions of the
inflaton field $\chi$.}
\label{slowrollparam}
\end{figure}

In order to let inflation last for a sufficient amount of time, the
time derivative of the Hubble parameter $H$ has to be sufficiently
small. As a result, the slow roll parameters
\beq
 \epsilon  =  -\frac{\dot{H}}{H^2},
\qquad
 \eta  = \epsilon-\frac{\dot{\epsilon}}{2\epsilon H}
= - \frac{\ddot{\chi}}{H\dot{\chi}},
\eeq
have to be much smaller than unity by modulus, leading to a negligible contribution
of the kinetic energy of the field during inflation. Using the Friedmann equations,
one can express the slow roll parameters in terms of the potential as
\beq
\label{slowparameters}
\epsilon = \frac{M_{P}^2}{2}\left[\frac{V'(\chi)}{V(\chi)} \right]^2
\quad
\mbox{and}
\quad
\eta = M_{P}^2\frac{V''(\chi)}{V(\chi)}.
\eeq
Finally, using Eq.~(\ref{rel_chi_phi}), the two parameters can be
written in terms of the field $\ph$,
\beq
\epsilon
& = &
\frac{4}{3}
\frac{1}{\left(1+k e^{2\ph}\right)\left(1-e^{2\ph}\right)^2},
\qquad
\label{parepsilon}
\\
\eta & = &
-\frac{4}{3}\left[\frac{e^{-2\ph}
\left(1-2e^{-2\ph} \right)
+ \frac{k}{2}
\left(3-5e^{-2\ph} \right)}{\left(1+k e^{2\ph} \right)^2
\left(1-e^{-2\ph} \right)^2}\right].
\qquad
\label{pareta}
\eeq
In the limit $k=0$ the slow roll parameters of the  $R+\al R^2$ model
are recovered. For small values of $k>0$, one can see that the slow
roll conditions are satisfied for large enough fields, while for $k<0$
there is a maximum value of the $\chi$ field where the slow roll
regime is valid: when $|k|\ll 1$. For  the values $k<0$ we have
slow roll for a wide range of the field, as shown in
Fig.~\ref{slowrollparam}.
\\

The number of inflationary $e$-folds $N$ in the Einstein frame is given by
\beq
 N(\chi)
 = \frac{1}{M_{P}^2}\int_{\chi_e}^{\chi}\,\,
 \frac{V(\chi)}{V'(\chi)}\,d \chi ,
\n{N}
\eeq
where $\chi_e$ corresponds to the end of inflation. In terms
of the field  $\varphi$ we get
\beq
\label{efold_number}
 N(\ph)=\left.\frac{3}{4}\Big[-2\ph +\left(1-k \right)e^{2\ph}
 +\frac{k}{2} e^{4\ph}\Big]\right|_{\ph_e}^{\ph}.
\eeq
The standard results for the $R+\al R^2$ model are recovered
for $k=0$. Assuming that $\chi_e \ll \chi$ (that is equivalent to
$\ph_e\ll\ph$ in the case of large field inflation), one can neglect the linear terms in
$\ph$ in Eq.~(\ref{efold_number}), which boils down to
\beq
\label{Nfolds_phi}
 N(\ph)\approx\frac{3}{4}\left[(1-k)e^{2\ph}+\frac{k}{2} e^{4\ph} \right].
\eeq
One can use this relation to derive the value $\ph_N$ of the field
$\ph$, corresponding to the instant when the universe expanded
by $N$ $e$-folds,
\beq
\label{CampoN}
 \ph_N \,\approx \, \frac{1}{2}
 \ln\left[1 - \frac{1}{k}
 \pm \sqrt{\Big(1-\frac{1}{k}\Big)^2
 +\frac{8N}{3k}}\right],
\eeq
where the positive sign has to be taken in order to recover the results for the $R+R^2$
model. Taking into account that $|k|\ll 1$ we can simplify this expression as
\begin{equation}
 \varphi_N\approx \frac{1}{2}\ln\left[-\frac{1}{k}+\sqrt{\frac{1}{k^2}+\frac{8N}{3k}}\right].
\end{equation}
When the slow roll conditions are satisfied, the amplitude of scalar (curvature)
and tensor perturbations can be written in terms of the potential $V(\chi)$ and its
derivatives at the moment when their physical wavelength $\lambda=p/a(t),~p=const$
crosses the Hubble radius $H^{-1}(t)$ during inflation. Here $a(t)$ is the
scale factor of an isotropic universe and $H(t)=\dot a(t)/a(t)$. The same
matching condition helps to express $N$ as a function of the present physical scale
$\lambda = a(t_0)/p$ where $t_0$ is the present moment. Then one gets the
standard expressions for the spectral index $n_s(p)$ of the power spectrum
of primordial curvature perturbations and the tensor-to-scalar ratio $r(p)$
in the leading order of the slow-roll approximation:
\begin{eqnarray}
 n_s-1 &=& -6\epsilon+2\eta=M_P^2\left(2\,\frac{V''}{V}-3\left(\frac{V'}{V}\right)^2\right),\\
r  &=& 16\epsilon=-8M_P^2 \left(\frac{V'}{V}\right)^2~.
\end{eqnarray}
Because of the conformal transformation between the Jordan and Einstein frames,
the same value of a perturbation as a function of $N$ and, finally, $p$ corresponds
to somewhat different physical scales in the Jordan and the Einstein frames. Since
the standards of length and time intervals are defined in Jordan
frame which can be considered as the physical one from the measurement point of
view,  the number of e-folds in the Jordan frame $N_J$ is more directly related to observations. However, difference between $N$ and $N_J$ is small, of the order of
the next correction to the slow-roll approximation  (less than a few percent for the
model in question), so we may neglect it in the leading order.

Using equations (\ref{parepsilon}) and (\ref{pareta}), we can obtain the following
analytical expressions for $n_s$ and $r$ as functions of $k$ and the number of
e-folds N:
  \begin{eqnarray}\label{ns-r}
   n_s-1&=&\frac{9k\left[18k^3+6k^2(Sq_{k,N}+12N-29)+8N(9-2Sq_{k,N})+40Sq_{k,N}-138\right]}
   {(Sq_{k,N}-3)^2(Sq_{k,N}+3k)^2}
  +
   \nonumber
\\
&&
\frac{9k^2\left[16N(Sq_{k,N}+4N-18)-52(Sq_{k,N}+261)\right]-54Sq_{k,N}+18}
{(Sq_{k,N}-3)^2(Sq_{k,N}+3k)^2},
\nonumber
\\
r&=&\frac{576 k^2}{(Sq_{k,N}-3)^2(Sq_{k,N}+3k)^2},
\end{eqnarray}
where we have defined $Sq_{k,N}=\sqrt{24kN+9(k-1)^2}$.\\
From equations (\ref{ns-r}), we can realize that when $k=0$, we
recover the predictions of the  $R+R^{2}$ model
\cite{Starobinsky:1983zz, mukhanov1981quantum},
\begin{eqnarray}
 n_s-1&\approx &-\frac{2}{N} \qquad r\approx \frac{12}{N^2},
\end{eqnarray}
and up to the order $kN$ there is no shift in $n_s-1$ and $r$, and their
corrections are of order $k/N$ or $k/N^2$, respectively.

\begin{figure}
\centering
\includegraphics[scale=1.2]{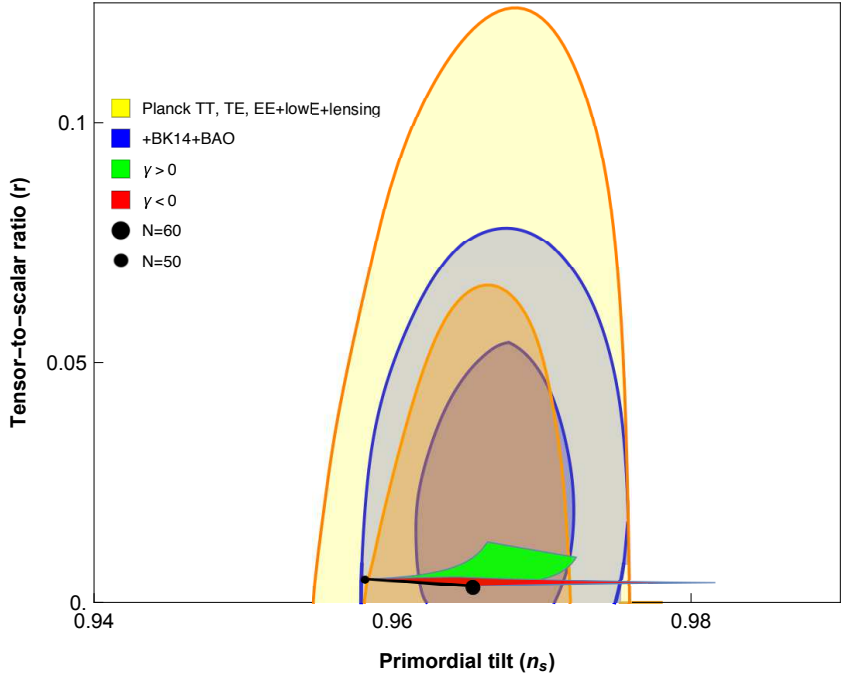}
\caption{Observational constraints set by the Planck collaboration
\cite{aghanim2018planck,Akrami:2018odb} on the scalar-to-tensor ratio $r$ and scalar
spectral index $n_s$, and the prediction for these quantities in the
$R + R^2 + R\Box R$ model.}
\label{constraintsplanck}
\end{figure}

The comparison of this inflationary model with the
observational constraints set by the Planck collaboration
\cite{aghanim2018planck,Akrami:2018odb}  is illustrated
in Fig.~\ref{constraintsplanck}.
The last Planck data constrain these quantities as
\beq
 n_s=0.9649\pm0.0042, \qquad r<0.10.
\eeq
In Fig.~\ref{constraintsplanck} we show the Planck constraints on the
values of $n_s$ and $r$, and the prediction for this quantities in the
$R+\al R^2+\ga R\Box R$ model. This figure shows the 68\% (dark
blue and dark yellow) and 95\% (light blue and light yellow) CL regions
for the measurements of $r$ and $n_s$, taking the combined data as
stated in the plot key, and the variation of the $R+\al R^2$ model
(black line) due to the inclusion of the $\ga$ term regarded as
a small correction for $k>0$ and $k<0$ in green and red, respectively.

Using a numerical routine, from Fig.~\ref{constraintsplanck} we have
found the maximum positive value of $k$ in order to keep the
predictions of the $R \Box R$ model inside the 68\% CL region,
concluding that, for $N=50$ and $N=55$ e-folds, any value of $k$
satisfying the perturbative condition $|k|\ll 1$ keeps the predictions
of the $R\Box R $ model inside the 68\% CL region. For the case
$N=60$ the maximum value of $k$ that satisfies this condition is
$k_{\rm max}\approx 0.30$.
\\
On the other hand, for negative values of $k$, Planck results
constraints the minimum value of $k$ to $-0.0060$, $-0.0052$
and $-0.0045$, when $N=50$, 55 and 60, respectively.
\\

\begin{figure}
\centering
\includegraphics[scale=1.2]{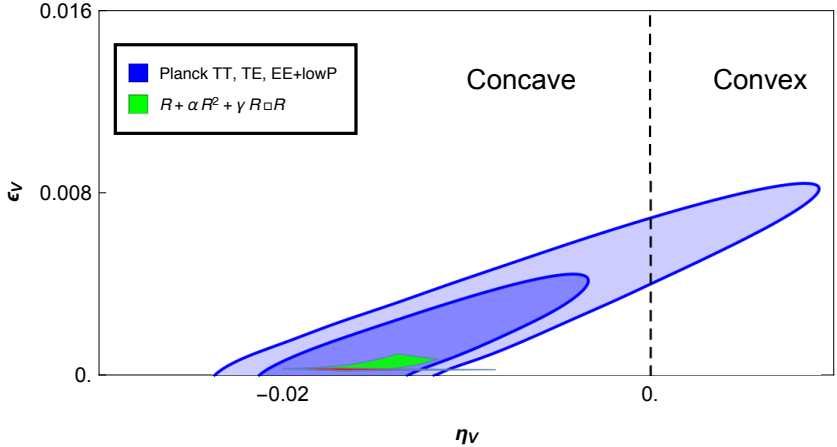}
\caption{Observational constraints on the slow roll parameters set by
Planck collaboration \cite{ade2016planck}, and the
prediction for these quantities in the $R+\al R^2+\ga R\Box R$.}
\label{slowrollplanck}
\end{figure}

Furthermore, Fig.~\ref{slowrollplanck} shows the Planck constraints
\cite{ade2016planck,Akrami:2018odb} for the relation between the slow roll parameters (\ref{slowparameters}), at the 68\% (dark blue) and 95\% CL (light blue). Also in this figure one can see the prediction of our present model (light green).
\\

Finally, one can check how the upper bound for the energy scale $M$
may change in the case of the $R+R^2+R\Box R$ model. Using a
procedure similar to the one presented in \cite{kaneda2010fourth}, we
use the expression (\ref{corrpot}) for the inflationary potential and
Eq.~(\ref{parepsilon}) for the first slow roll parameter, written in
terms of the number of $e$-folds $N$, from which we get the
curvature perturbations
\beq
\Delta_{\mathcal{R}}^2=\frac{1}{M_{P}^2}\sqrt{\frac{8V}{3\epsilon}},
\label{Delta}
\eeq
and make a comparison with the last Planck results \cite{aghanim2018planck}.
For the case of 60 $e$-folds and using the value of $k$ which assures
that the correction due to the $R\Box R$ term lies inside the 68\% CL
regions of both figures \ref{constraintsplanck} and \ref{slowrollplanck}
(hence $k\approx 0.30$), the maximum value of $M$ would be
\beq
M_{\rm max}=(2.21\pm0.01)\times 10^{-5}M_P,
\eeq
which is consistent with the result of \cite{Starobinsky:1983zz}.

\section{Conclusions}
\label{s5}

We consider the extension of the
$R+R^2$ inflationary model by adding a small perturbation of
the form $R\Box R$. Treating terms with higher than four
derivatives as small perturbations has been suggested as a
general approach to deal with higher derivative terms in quantum gravity,
and following this approach we included, for the first time, the $R^2$ term
into the main, non-perturbative part
of the action. We have shown that there is no $kN$ relative correction to the
main terms in $n_s-1$ and $r$ when the $R\Box R$ term is treated as
a small perturbation to the Starobinsky model.
First of all, this means that the presence of this term does not
increase the amount of degrees of freedom, and one has to perform
the mapping of the modified gravity theory to the scalar-tensor
model with only one effective scalar field. This approach opens the way
for a simple and explicit analysis of observational constraints
on extra terms. The case which we considered here provides
an especially simple mapping procedure, but there is a good chance
that the same result can be achieved for other higher derivative
extensions of general relativity.

\appendix
\section{Action in terms of scalar fields}\label{ap1}

Let us briefly review the standard treatment of the model under
discussion, which implies using two scalar fields. For this end
we write the action (\ref{R-action}) as \cite{chiba2005generalized}
(see also \cite{PhysRevD.83.084028})
\beq\label{action1_apen}
S
&=&
\frac{M_P^2}{2}\int d^4x\sqrt{-g}\left[F(\phi_1,\phi_2)
+F_1(R-\phi_1)+F_2(\Box R-\phi_2) \right],
\eeq
where $F(\phi_1,\phi_2)=\phi_1+\alpha\phi_1^2+\gamma\phi_1\phi_2$,
$F_1=\frac{\partial F}{\partial \phi_1}$, and
$F_2=\frac{\partial F}{\partial \phi_2}$, in such a way that the action
takes the form
\beq
 S
 &=&
 \frac{M_P^2}{2}\int d^4x\sqrt{-g}\left[\phi_1+\alpha \phi_1^2
 +\gamma \phi_1\phi_2 +
 (1+2\alpha \phi_1+\gamma \phi_2)(R-\phi_1)
 +\gamma\phi_1(\Box R -\phi_2)\right]
 \nonumber
 \\
 &=&\frac{M_P^2}{2}\int d^4x\sqrt{-g}\left[(1+2\alpha \phi_1+\gamma\phi_2)R-\alpha\phi_1^2 +
 \gamma\phi_1\Box R-\gamma\phi_1\phi_2\right].
\end{eqnarray}
In order to eliminate the term with $\Box R$, we integrate by
parts \cite{wands1994extended},
\beq
 \int d^4x\sqrt{-g}\phi_1\Box R&=&
 -\int d^4x\sqrt{-g}\nabla^\mu \phi_1\nabla_\mu R
 =\int d^4x\sqrt{-g}R\Box \phi_1,
\end{eqnarray}
and arrive at the following expression for the action
\beq
S
&=&
\frac{M_P^2}{2}\int d^4x\sqrt{-g}\left[(1+2\alpha \phi_1+\gamma\phi_2+\gamma\Box\phi_1)R -
\alpha\phi_1^2-\gamma\phi_1\phi_2 \right].
\end{eqnarray}
Defining
$\,\sigma=1+2\alpha\phi_1+\gamma\phi_2+\gamma\Box\phi_1$,
the action can be written in terms of two scalar fields in the form
\beq
\label{scoupled}
S  =  \frac{M_P^2}{2}\int d^4x\sqrt{-g}\left[\sigma R+\gamma\phi_1\Box\phi_1-U(\phi_1,\sigma)\right],
\\
\mbox{where} \quad
U(\phi_1,\sigma)=\phi_1(\sigma-1)-\alpha\phi_1^2.
\nn
\eeq

From Eq.~(\ref{scoupled}) one can see that setting $\ga=0$,
we recover the $R+\alpha R^2$ case \cite{Starobinsky:1980te}.

In order to  analyze the theory with two scalar degrees of freedom,
we have to perform the conformal transformation of the metric,
$\bar{g}_{\mu\nu}=e^{2\varphi}g_{\mu\nu}$ \
\cite{dabrowski2009conformal,carneiro2004useful},
\begin{eqnarray}
 S
 &=&
 \frac{M_P^2}{2}
 \int d^4x\sqrt{-\bar{g}} e^{-4\ph}
 \left\{
 \si e^{2\varphi} \left[
 \bar{R} - 6 (\bar{\nabla} \ph)^2  - 6 \bar{\Box}\ph) \right]
 \right.
 \nonumber
 \\
 &+&\left.\gamma\phi_1e^{2\varphi}\left(\bar{\Box}\phi_1
 -2\bar{\nabla}^{\mu}\varphi\bar{\nabla}_\mu \phi_1\right)
 -U(\phi_1,\sigma)\right\}.
\end{eqnarray}
In this case, taking $\sigma=e^{2\varphi}$, it is straightforward
to obtain the action in the Einstein frame (\ref{sfinal}).

One can show that the fields $\phi_1$ and $\varphi$ satisfy the
equations of motion
\begin{eqnarray}
\label{eqmovphi1sigma}
 & & \bar{\Box}\phi_1
 -2\bar{\nabla}^{\mu}\varphi \bar{\nabla}_{\mu}\phi_1
 - \frac{e^{-2\varphi}}{2\gamma}\left[e^{2\varphi}
 -1-2\alpha\phi_1 \right]=0,
 \nonumber
 \\
&&
\Box \phi_1+\frac{\alpha}{\gamma}\phi_1
= - \frac{1}{2\gamma}(1-e^{2\varphi}),
\end{eqnarray}
which are consistent with the relation between the original fields
$\Box \phi_1=\phi_2$ and reproduces the results of~\cite{Gottlober:1989ww}.

\subsection{Weak field approximation}
The Einstein tensor can be found from equation (\ref{sfinal}),
taking the variation with respect to  
$g^{\mu\nu}$,
\begin{eqnarray}
\bar{G}_{\mu\nu}
&=&
\ga e^{-2\ph}
\Big(
\bar{\nabla}^{\mu}\phi_1\bar{\nabla}_{\nu}\phi_1 -\frac{1}{2}\bar{g}_{\mu\nu}\bar{\nabla}^{\la}\phi_1
 \bar{\nabla}_{\la}\phi_1\Big)
 + 6 \Big(\bar{\nabla}_{\mu}\ph\bar{\nabla}_{\nu}\ph -\frac{1}{2}\bar{g}_{\mu\nu}\bar{\nabla}^{\la}\ph
 \bar{\nabla}_{\la}\ph\Big)
 \nonumber
 \\
 &+&
 \frac12\,\bar{g}_{\mu\nu}e^{-4\varphi}
 \left[\al\phi_1^2+(1-e^{2\ph})\phi_1 \right].
\end{eqnarray}
The last expression is consistent with the Einstein tensor
obtained for the first time in \cite{Gottlober:1989ww}, where
the weak field approximation was worked out, showing that the
action in this case is given by
\begin{eqnarray}
S_{wf}
 &\approx&
\frac{M_P^2}{2}\int d^4x\sqrt{-\bar{g}}
\Big[
\bar{R}-6\bar{\na}^{\la}\ph\bar{\na}_{\la}\ph
- \ga\bar{\na}^{\la}\phi_1\bar{\na}_{\la}\phi_1
 - 2\phi_1\ph+\al\phi_1^2\Big].
\end{eqnarray}
The mixed term involving $\phi_1$ and $\ph$ can be
removed by performing a rotation of these fields, leading to the identification of
scalar fields that can be tachyonic or physical, depending
on the $\al$ and $\ga$ parameters \cite{berkin1990effects}.
In order to have a stable Minkowski space it is necessary that
$\ga<0$, within this approximation. Let us note that this
condition does not apply within our approach to the problem.

\subsection{Simplifying Ansatz \texorpdfstring{$\Box R=r_1 R+r_2$}{boxR}}

In this section we consider the simplifying Ansatz
$\Box R=r_1 R+r_2$ proposed in
\cite{craps2014cosmological}, and later on used in
\cite{koshelev2016occurrence} to analyze non-local modifications
of gravity with general form factors depending on the D'Alambertian
operator $\Box$ applied to the Riemann, Ricci tensors and to the
Ricci scalar.

Our main goal is to show that the application of this simplifying
Ansatz to the action (\ref{sfinal}), under certain conditions for the
parameters $\alpha$ and $\gamma$, reproduces the results obtained
in \cite{craps2014cosmological,koshelev2016occurrence} for this
particular case.

As far as we are not considering the cosmological constant term in
the action, the $r_2$ contribution vanishes, and the ansatz becomes
\begin{equation}
\label{ansatz}
\Box \phi_1=r_1\phi_1.
\end{equation}
The action given by (\ref{sfinal}), can be written as
\beq
 S
 =
 \frac{M_P^2}{2}\int d^4x\sqrt{-\bar{g}}
 \left[\bar{R}-6({\bar{\nabla}\varphi})^2
 +\gamma\phi_1e^{-2\varphi}\left(\bar{\Box}\phi_1-2\bar{\nabla}^{\mu}\ph \bar{\nabla}_{\mu}\phi_1\right)
-U(\phi_1,\varphi)\right],
\n{1}
\eeq
or equivalently, using Eq.~(\ref{ansatz}), as
\beq
S
&=&
\frac{M_P^2}{2}\int d^4x\sqrt{-\bar{g}}\left[\bar{R}
-6({\bar{\nabla}\varphi})^2+\ga r_1\phi_1^2e^{-4\ph}\right.
\left.- U(\phi_1,\varphi)\right].
\n{2}
\eeq

Finally, using the relation between the fields $\si=e^{2\ph}$
and $\phi_1$,  we arrive at the one-scalar representation
\beq
S &=&
\frac{M_P^2}{2}\int d^4x\sqrt{-\bar{g}}\left[\bar{R}
-6({\bar{\nabla}\varphi})^2 -U(\ph)\right],
\n{3}
\eeq
with the potential
\beq
\label{pot_ansatz}
 V(\ph)=\frac{U(\ph)}{2}=\frac{1}{8(\alpha+\ga r_1)}(1-e^{-2\ph})^2.
\eeq

From (\ref{Potencialcero}) and (\ref{pot_ansatz}), it is clear that
assuming that the scalar field $\ph$ evolves within the slow roll
approximation, the scalar spectral index $n_s$ and
the tensor-to-scalar ratio $r$, which depend on the potential, are
exactly the same as obtained in $R^2$ inflation, as was previously
shown by \cite{craps2014cosmological} for the model $R+R^n$, and
after them,  in the case of a non-local framework. There is no change
in the tensor-to-scalar ratio $r$, because the Weyl term $C^2$ is
excluded and the non-local operator $F_C(\Box)$ is absent
\cite{koshelev2016occurrence} in our analysis.
It is important to remark that, as was shown in Refs.
\cite{koshelev2016occurrence,craps2014cosmological},
any solution of the $R+R^2$ theory satisfying the
ansatz (\ref{ansatz}), is also a solution of the  $R+R^2+R\Box R$
theory, but this is not the whole solution to the theory, as there could
be solutions to $R+R^2+R\Box R$ which do not satisfy the simplifying
ansatz (\ref{ansatz}). On the other hand, if we assume that inflation
is just as in the $R+R^2$ model, we can take the $R\Box R$ term as a
small perturbation, in the same spirit in which Ref. \cite{Simon:1990ic}
analyzes higher order terms as corrections to the Einstein-Hilbert action,
leading to the determination of new solutions of the corrected theory.
In our case, these solutions are independent of the simplifying ansatz
\eqref{ansatz}, but give the same results when this assumption is
taken as a particular case.

\acknowledgments
A.R.R.C. is supported in part by CAPES - Code 001.
I.Sh. is grateful to CNPq (grant 303893/2014-1) and 
FAPEMIG (grant APQ-01205-16) for partial support.
F.S. is grateful to CNPq for partial support.
A.A.S. is supported by the RSF grant 16-12-10441.

\bibliographystyle{JHEP}
\bibliography{refs-Ana-v3g}

\providecommand{\href}[2]{#2}\begingroup\raggedright\begin{thebibliography}{10}

\bibitem{Utiyama:1962sn}
R.~Utiyama and B.~S. DeWitt, {\it {Renormalization of a classical gravitational
  field interacting with quantized matter fields}},  {\em J. Math. Phys.} {\bf
  3} (1962) 608--618.

\bibitem{Zeldovich:1971mw}
{\relax Ya}.~B. Zeldovich and A.~A. Starobinsky, {\it {Particle production and
  vacuum polarization in an anisotropic gravitational field}},  {\em Sov. Phys.
  JETP} {\bf 34} (1972) 1159--1166. [Zh. Eksp. Teor. Fiz. 61, 2161 (1971)].

\bibitem{Shapiro:2008sf}
I.~L. Shapiro, {\it {Effective Action of Vacuum: Semiclassical Approach}},
  {\em Class. Quant. Grav.} {\bf 25} (2008) 103001,
  [\href{http://arxiv.org/abs/0801.0216}{{\tt arXiv:0801.0216}}].

\bibitem{Stelle:1976gc}
K.~S. Stelle, {\it {Renormalization of Higher Derivative Quantum Gravity}},
  {\em Phys. Rev.} {\bf D16} (1977) 953--969.

\bibitem{Salam:1978fd}
A.~Salam and J.~A. Strathdee, {\it {Remarks on High-energy Stability and
  Renormalizability of Gravity Theory}},  {\em Phys. Rev.} {\bf D18} (1978)
  4480.

\bibitem{Tomboulis:1977jk}
E.~Tomboulis, {\it {1/N Expansion and Renormalization in Quantum Gravity}},
  {\em Phys. Lett.} {\bf 70B} (1977) 361--364.

\bibitem{Tomboulis:1980bs}
E.~Tomboulis, {\it {Renormalizability and Asymptotic Freedom in Quantum
  Gravity}},  {\em Phys. Lett.} {\bf 97B} (1980) 77--80.

\bibitem{Tomboulis:1983sw}
E.~T. Tomboulis, {\it {Unitarity in Higher Derivative Quantum Gravity}},  {\em
  Phys. Rev. Lett.} {\bf 52} (1984) 1173.

\bibitem{Antoniadis:1986tu}
I.~Antoniadis and E.~T. Tomboulis, {\it {Gauge Invariance and Unitarity in
  Higher Derivative Quantum Gravity}},  {\em Phys. Rev.} {\bf D33} (1986) 2756.

\bibitem{Johnston:1987ue}
D.~A. Johnston, {\it {Sedentary Ghost Poles in Higher Derivative Gravity}},
  {\em Nucl. Phys.} {\bf B297} (1988) 721--732.

\bibitem{Asorey:1996hz}
M.~Asorey, J.~L. Lopez, and I.~L. Shapiro, {\it {Some remarks on high
  derivative quantum gravity}},  {\em Int. J. Mod. Phys.} {\bf A12} (1997)
  5711--5734, [\href{http://arxiv.org/abs/hep-th/9610006}{{\tt
  hep-th/9610006}}].

\bibitem{Modesto:2015ozb}
L.~Modesto and I.~L. Shapiro, {\it {Superrenormalizable quantum gravity with
  complex ghosts}},  {\em Phys. Lett.} {\bf B755} (2016) 279--284,
  [\href{http://arxiv.org/abs/1512.07600}{{\tt arXiv:1512.07600}}].

\bibitem{Simon:1990ic}
J.~Z. Simon, {\it {Higher Derivative Lagrangians, Nonlocality, Problems and
  Solutions}},  {\em Phys. Rev.} {\bf D41} (1990) 3720.

\bibitem{Starobinsky:1980te}
A.~A. Starobinsky, {\it {A New Type of Isotropic Cosmological Models Without
  Singularity}},  {\em Phys. Lett.} {\bf B91} (1980) 99--102. [,771(1980)].

\bibitem{Simon:1991bm}
J.~Z. Simon, {\it {No Starobinsky inflation from selfconsistent semiclassical
  gravity}},  {\em Phys. Rev.} {\bf D45} (1992) 1953--1960.

\bibitem{Starobinsky:1983zz}
A.~A. Starobinsky, {\it {The Perturbation Spectrum Evolving from a Nonsingular
  Initially De-Sitter Cosmology and the Microwave Background Anisotropy}},
  {\em Sov. Astron. Lett.} {\bf 9} (1983) 302.

\bibitem{kaneda2010fourth}
S.~Kaneda, S.~V. Ketov, and N.~Watanabe, {\it Fourth-order gravity as the
  inflationary model revisited},  {\em Modern Physics Letters A} {\bf 25}
  (2010), no.~32 2753--2762.

\bibitem{Netto:2015cba}
T.~d.~P. Netto, A.~M. Pelinson, I.~L. Shapiro, and A.~A. Starobinsky, {\it
  {From stable to unstable anomaly-induced inflation}},  {\em Eur. Phys. J.}
  {\bf C76} (2016), no.~10 544, [\href{http://arxiv.org/abs/1509.08882}{{\tt
  arXiv:1509.08882}}].

\bibitem{Gottlober:1989ww}
S.~Gottlober, H.~J. Schmidt, and A.~A. Starobinsky, {\it {Sixth Order Gravity
  and Conformal Transformations}},  {\em Class. Quant. Grav.} {\bf 7} (1990)
  893.

\bibitem{berkin1990effects}
A.~L. Berkin and K.-i. Maeda, {\it Effects of ${R^3}$ and ${R\Box R}$ terms on
  ${R^2}$ inflation},  {\em Physics Letters B} {\bf 245} (1990), no.~3-4
  348--354.

\bibitem{Amendola:1993bg}
L.~Amendola, A.~Battaglia~Mayer, S.~Capozziello, F.~Occhionero, S.~Gottlober,
  V.~Muller, and H.~J. Schmidt, {\it {Generalized sixth order gravity and
  inflation}},  {\em Class. Quant. Grav.} {\bf 10} (1993) L43--L47.

\bibitem{chiba2005generalized}
T.~Chiba, {\it Generalized gravity and a ghost},  {\em Journal of Cosmology and
  Astroparticle Physics} {\bf 2005} (2005), no.~03 008.

\bibitem{Cuzinatto:2013pva}
R.~R. Cuzinatto, C.~A.~M. de~Melo, L.~G. Medeiros, and P.~J. Pompeia, {\it
  {Observational constraints on a phenomenological $f\left( R,\partial R\right)
  $-model}},  {\em Gen. Rel. Grav.} {\bf 47} (2015), no.~3 29,
  [\href{http://arxiv.org/abs/1311.7312}{{\tt arXiv:1311.7312}}].

\bibitem{craps2014cosmological}
B.~Craps, T.~De~Jonckheere, and A.~S. Koshelev, {\it Cosmological perturbations
  in non-local higher-derivative gravity},  {\em Journal of Cosmology and
  Astroparticle Physics} {\bf 2014} (2014), no.~11 022.

\bibitem{koshelev2016occurrence}
A.~S. Koshelev, L.~Modesto, L.~Rachwal, and A.~A. Starobinsky, {\it Occurrence
  of exact r 2 inflation in non-local uv-complete gravity},  {\em Journal of
  High Energy Physics} {\bf 2016} (2016), no.~11 67.

\bibitem{Carloni:2018yoz}
S.~Carloni, J.~L. Rosa, and J.~P.~S. Lemos, {\it {Cosmology of $f(R, \square
  R)$ gravity}},  \href{http://arxiv.org/abs/1808.07316}{{\tt
  arXiv:1808.07316}}.

\bibitem{Maroto:1997aw}
A.~L. Maroto and I.~L. Shapiro, {\it {On the inflationary solutions in higher
  derivative gravity with dilaton field}},  {\em Phys. Lett.} {\bf B414} (1997)
  34--44, [\href{http://arxiv.org/abs/hep-th/9706179}{{\tt hep-th/9706179}}].

\bibitem{linde1984inflationary}
A.~D. Linde, {\it The inflationary universe},  {\em Reports on Progress in
  Physics} {\bf 47} (1984), no.~8 925.

\bibitem{olive1990inflation}
K.~A. Olive, {\it Inflation},  {\em Physics Reports} {\bf 190} (1990), no.~6
  307--403.

\bibitem{bassett2006inflation}
B.~A. Bassett, S.~Tsujikawa, and D.~Wands, {\it Inflation dynamics and
  reheating},  {\em Reviews of Modern Physics} {\bf 78} (2006), no.~2 537.

\bibitem{Fischetti:1979ue}
M.~V. Fischetti, J.~B. Hartle, and B.~L. Hu, {\it {Quantum Effects in the Early
  Universe. 1. Influence of Trace Anomalies on Homogeneous, Isotropic,
  Classical Geometries}},  {\em Phys. Rev.} {\bf D20} (1979) 1757--1771.

\bibitem{Fabris:2000gz}
J.~C. Fabris, A.~M. Pelinson, and I.~L. Shapiro, {\it {On the gravitational
  waves on the background of anomaly-induced inflation}},  {\em Nucl. Phys.}
  {\bf B597} (2001) 539--560, [\href{http://arxiv.org/abs/hep-th/0009197}{{\tt
  hep-th/0009197}}]. [Erratum: Nucl. Phys.B602,644(2001)].

\bibitem{Fabris:2003gp}
J.~C. Fabris, A.~M. Pelinson, I.~L. Shapiro, and F.~I. Takakura, {\it
  {Gravitational waves in an anomaly induced inflation}},  {\em Nucl. Phys.
  Proc. Suppl.} {\bf 127} (2004) 159--161,
  [\href{http://arxiv.org/abs/hep-ph/0311309}{{\tt hep-ph/0311309}}].
  [,159(2003)].

\bibitem{Akrami:2018vks}
{\bf Planck} Collaboration, Y.~Akrami et~al., {\it {Planck 2018 results. I.
  Overview and the cosmological legacy of Planck}},
  \href{http://arxiv.org/abs/1807.06205}{{\tt arXiv:1807.06205}}.

\bibitem{Akrami:2018odb}
{\bf Planck} Collaboration, Y.~Akrami et~al., {\it {Planck 2018 results. X.
  Constraints on inflation}},  \href{http://arxiv.org/abs/1807.06211}{{\tt
  arXiv:1807.06211}}.

\bibitem{aghanim2018planck}
{\bf Planck} Collaboration, N.~Aghanim et~al., {\it Planck 2018 results. vi.
  cosmological parameters},  \href{http://arxiv.org/abs/1807.06209}{{\tt
  arXiv:1807.06209}}.

\bibitem{PhysRevD.83.084028}
D.~C. Rodrigues, F.~d.~O. Salles, I.~L. Shapiro, and A.~A. Starobinsky, {\it
  Auxiliary fields representation for modified gravity models},  {\em Phys.
  Rev. D} {\bf 83} (Apr, 2011) 084028.

\bibitem{huang2014polynomial}
Q.-G. Huang, {\it A polynomial f (r) inflation model},  {\em Journal of
  Cosmology and Astroparticle Physics} {\bf 2014} (2014), no.~02 035.

\bibitem{Accioly:2016qeb}
A.~Accioly, B.~L. Giacchini, and I.~L. Shapiro, {\it {Low-energy effects in a
  higher-derivative gravity model with real and complex massive poles}},  {\em
  Phys. Rev.} {\bf D96} (2017), no.~10 104004,
  [\href{http://arxiv.org/abs/1610.05260}{{\tt arXiv:1610.05260}}].

\bibitem{mukhanov1998density}
V.~F. Mukhanov and P.~J. Steinhardt, {\it Density perturbations in multifield
  inflationary models},  {\em Physics Letters B} {\bf 422} (1998), no.~1-4
  52--60.

\bibitem{garcia1995constraints}
J.~Garcia-Bellido and D.~Wands, {\it Constraints from inflation on
  scalar-tensor gravity theories},  {\em Physical Review D} {\bf 52} (1995),
  no.~12 6739.

\bibitem{garcia1996metric}
J.~Garcia-Bellido and D.~Wands, {\it Metric perturbations in two-field
  inflation},  {\em Physical Review D} {\bf 53} (1996), no.~10 5437.

\bibitem{di2005slow}
F.~Di~Marco and F.~Finelli, {\it Slow-roll inflation for generalized two-field
  lagrangians},  {\em Physical Review D} {\bf 71} (2005), no.~12 123502.

\bibitem{wang2010note}
T.~Wang, {\it Note on non-gaussianities in two-field inflation},  {\em Physical
  Review D} {\bf 82} (2010), no.~12 123515.

\bibitem{ji2009curvature}
X.~Ji and T.~Wang, {\it Curvature and entropy perturbations in generalized
  gravity},  {\em Physical Review D} {\bf 79} (2009), no.~10 103525.

\bibitem{lalak2007curvature}
Z.~Lalak, D.~Langlois, S.~Pokorski, and K.~Turzy{\'n}ski, {\it Curvature and
  isocurvature perturbations in two-field inflation},  {\em Journal of
  Cosmology and Astroparticle Physics} {\bf 2007} (2007), no.~07 014.

\bibitem{wang2018noncanonical}
Y.-C. Wang and T.~Wang, {\it Noncanonical two-field inflation to order $\xi$
  2},  {\em International Journal of Modern Physics D} {\bf 27} (2018), no.~03
  1850026.

\bibitem{Starobinsky:1981vz}
A.~A. Starobinsky, {\it {Nonsingular model of the Universe with the
  quantum-gravitational de Sitter stage and its observational consequences}},
  in {\em {Second Seminar on Quantum Gravity Moscow, USSR, October 13-15,
  1981}}, pp.~103--128, 1981.

\bibitem{kofman1994reheating}
L.~Kofman, A.~Linde, and A.~A. Starobinsky, {\it Reheating after inflation},
  {\em Physical Review Letters} {\bf 73} (1994), no.~24 3195.

\bibitem{kofman1997towards}
L.~Kofman, A.~Linde, and A.~A. Starobinsky, {\it Towards the theory of
  reheating after inflation},  {\em Physical Review D} {\bf 56} (1997), no.~6
  3258.

\bibitem{mukhanov1981quantum}
V.~F. Mukhanov and G.~Chibisov, {\it Quantum fluctuations and a nonsingular
  universe},  {\em JETP Lett.} {\bf 33} (1981) 532--535.

\bibitem{ade2016planck}
P.~Ade et~al., {\it Planck 2015 results-xx. constraints on inflation},  {\em
  Astronomy \& Astrophysics} {\bf 594} (2016) A20.

\bibitem{wands1994extended}
D.~Wands, {\it Extended gravity theories and the einstein--hilbert action},
  {\em Classical and Quantum Gravity} {\bf 11} (1994), no.~1 269.

\bibitem{dabrowski2009conformal}
M.~P. Dabrowski, J.~Garecki, and D.~B. Blaschke, {\it Conformal transformations
  and conformal invariance in gravitation},  {\em Annalen der Physik} {\bf 18}
  (2009), no.~1 13--32.

\bibitem{carneiro2004useful}
D.~F. Carneiro, E.~A. Freiras, B.~Gon{\c{c}}alves, A.~G. de~Lima, and
  I.~Shapiro, {\it On useful conformal tranformations in general relativity},
  {\em Gravitation and Cosmology} {\bf 10} (2004), no.~4 305--312.

\end{thebibliography}\endgroup
\end{document}